\documentclass[12pt]{article}
\usepackage{float}

\usepackage{sbc-template}
\usepackage{amsmath}
\usepackage[utf8]{inputenc}
\usepackage{graphicx}
\usepackage{algorithm}
\usepackage[noend]{algpseudocode}
\usepackage{epsf}
\usepackage{url}
\usepackage{comment}

\usepackage[dvips]{epsfig}

\title{Q-caching: an integrated reinforcement-learning approach for caching and routing in  information-centric networks}

\author{
Wouter Caarls, Eduardo Hargreaves, Daniel S. Menasch\'e}

\address{  Universidade Federal do Rio de Janeiro (UFRJ)\\
  Av. Athos da Silveira Ramos, 149 - Blc C - 21.941-909\\
  Rio de Janeiro, Brasil
\email{wouter@caarls.org,eduardo.hargreaves@ppgi.ufrj.br,sadoc@dcc.ufrj.br}}

\begin{document}
\maketitle

\begin{abstract}
Content delivery, such as  video streaming,  is one of the most prevalent Internet applications.   Although very popular, the continuous growth of such applications poses novel performance and scalability challenges.  
Information-centric networks put content at the center, and propose novel solutions to such challenges but also pose new questions on the interface between caching and routing.  In this paper, building on top of Q-routing we propose a caching strategy, namely Q-caching,  which leverages information that is already collected by the routing algorithm.  Q-caching promotes content diversity in the network, reducing the load at custodians and average download times for clients.  
In stylized topologies, we show that the gains of Q-caching against state-of-the-art algorithms are significant.    
We then consider  the RNP topology, and show that Q-caching performance is more flexible while  competitive when 
compared against existing algorithms.
\end{abstract}




\section{Introduction}


Information-centric networking (ICN) is a new networking paradigm that shifts the basic network service semantics from \textit{delivering the packet to a given destination address} to \textit{ fetching data identified by a given name} \cite{Afanasyev2014a,jacobson2009networking}.  In the ICN paradigm, routers are equipped with caches, with the aim of increasing network performance and scalability.  As the network service is content-aware, users can opportunistically download replicas of the content which are closer to the requesters.  

In traditional IP networks when there is a change in the network, routing protocols need to exchange routing updates in order to maintain the topology consistent. Thus, IP routing protocols need to converge fast in order to keep packet delivery after network changes. Fortunately, unless a failure occurs, IP networks are typically static in short timescales.

On the other hand, ICN is a extremely dynamic because it is a highly distributed caching network, with content location and availability changing over time. In this case, even a static network will have dynamic properties, as the location of content may change due to caching. These differences made \cite{yi2014role} rethink the role of routing in ICN.

Q-routing~\cite{NIPS1993_770} was proposed to address the problem of packet routing in dynamically changing networks. It uses single-hop delays to continually estimate the cost to retrieve a packet, called cost-to-go, essentially implementing an asynchronous version of the Bellman-Ford shortest paths algorithm using only local information. It is asynchronous because each router uses only local informational without any route update message.

In recent work~\cite{chiocchetti2013inform}, Q-routing has been extended to work in information-centric networks, by estimating not the best routing decision for a particular destination, but for a particular piece of content.

In this paper, we propose a reinforcement learning based solution to the joint problem of content placement and routing which naturally embraces the notion of utility for the two decisions motivated by the fact that 
leveraging advantages of ubiquitous caching requires coordination between content placement (insertion and eviction decisions) and content search (routing decisions) \cite{rossini2014coupling}. In particular, we exploit the fact that Q-routing computes the \emph{cost-to-go }and use it to make not only routing but also caching decisions in a weighted least-frequently used (WLFU) manner, evicting the item with the \emph{minimum expected cost} (MEC) to retrieve. Then, Q-routing and MEC are combined in order to efficiently route requests and store content with the goal of minimizing the download time experienced by users. The resulting combination is referred to as Q-caching.

With this kind of approach, the routing and caching decisions therefore influence each other, as caching an item will cause the routes to change, and changing a route changes the requests seen by other nodes in the network, changing their caches. It is therefore important for the two to be well-coordinated, which we realize by basing them on the same information. 

 There have been separate efforts to incorporate the concept of utility for placement (e.g., web proxy caching~\cite{Wooster:1997:PCE:283554.283259}) and search decisions (e.g., Q-routing~\cite{NIPS1993_770}).
Nonetheless, to the best of our knowledge none of the prior works accounted for the notion of utility in an integrated manner for placement and routing decisions.

The paper is organized as follows. 
First, we  introduce Q-routing for information-centric networking in Section~\ref{sec:qrouting}.
 Then, we  propose our cost-to-go based caching strategy in Section~\ref{sec:qcaching}.  
 In Sections~\ref{sec:local1} and~\ref{sec:features} we introduce the local cost minimization problem and some 
 of the key features of Q-caching.
 Section~\ref{workload} describes the workload used in our trace-driven experiments.   
 We will continue by presenting our preliminary results, related work and discussions in Sections~\ref{sec:experiments},~\ref{sec:related} and~\ref{sec:discussion}, and drawing conclusions in Section~\ref{sec:conclusion}.

\section{Q-routing}
\label{sec:qrouting}

In Q-routing~\cite{NIPS1993_770}, the cost-to-go is estimated locally and asynchronously using reinforcement learning~\cite{sutton98}. It uses a tabular \emph{value function} $Q_x(d, y)$ which stores, 
for every node $x$, the cost to reach destination $d$ when routing through immediate hop $y$. In the original paper $d$ is a destination \emph{node}, but in ICNs it is a piece of information.

The Q table is estimated on-line via the Q-learning update rule:

\begin{equation}
\label{eq:update}
Q_x(d, y) \gets Q_x(d, y) + \alpha\left(t + \left(\min_{y'\in\mathcal{N}(y)}Q_y(d, y') \right) - Q_x(d, y)\right)
\end{equation}
where $\alpha$ is the parameter of an exponential moving average filter, $\mathcal{N}(y)$ is the set of immediate neighbors of node $y$ and $t$ is the cost (e.g., delay) between $x$ and $y$. The minimum operation in Equation~\ref{eq:update} is performed by node $y$ when it receives the request from node $x$, and returned as part of the acknowledgement.

When routing a request, the Q table is consulted to find the neighbor y with the lowest cost-to-go for a particular piece of information $d$. Although in general some form of \emph{exploration} is necessary to find optimal routes (by routing to nodes which are not currently thought to provide the lowest cost), in practice the algorithm has shown to perform well~\cite{NIPS1993_770}. Alternatively, a separate \emph{exploration phase} \cite{chiocchetti2013inform} may be used to first estimate the $Q$ values.

Note that  Equation~\ref{eq:update} estimates the expected cost incurred by node $x$ to fech content $d$  through its neighbor $y$. 
Let $u(x,d,y)$ be the instantaneous local utility associated to  sending a request for content $d$ from node $x$ to its neighbor $y$. 
In general, any utility function  may be used in our estimates,  by letting  $t=-u(x,d,y)$.  For example, some content may be treated preferentially, or some nodes may be avoided.

\section{Q-caching}
\label{sec:qcaching}

In previous work, Q-routing was used to find content in an information centric network using least-recently used (LRU) caching~\cite{chiocchetti2013inform}. LRU is a ubiquitous caching strategy due to its fast and easy implementation, combined with good performance. However, the availability of the cost-to-go allows us to make better caching decisions than this simple heuristic. In particular, LRU does not take into account the downstream availability of the content.

If we wish to minimize the path delay for a given request, it makes sense to cache the items that are most difficult to obtain, i.e. have the highest cost-to-go $Q_x(d)$. 
Given a request distribution $r_x(d)$ that specifies the probability of a request for information $d$ arriving at node $x$, we wish to minimize the expected cost:
\begin{equation}
\sum_d r_x(d)Q_x(d) = \sum_d r_x(d)\min_{y\in\mathcal{N}(x) }Q_x(d, y) 
\end{equation}
This is achieved by sorting the content according to this expected cost and caching the items with the highest values. If $r_x(d)$ is estimated by request counting, this amounts to a weighted least-frequently used (WLFU) caching policy, where the weight is the cost to obtain the content~\cite{Wooster:1997:PCE:283554.283259}.

\begin{algorithm}
\caption{Q-caching}\label{alg:qrouting}
\begin{algorithmic}[1]
\State $Q \gets 0$, $\hat{r} \gets 0$ 
\State $C \gets \emptyset$
\Procedure{Q-caching}{x,d}
\State $\hat{r}_x(d) \gets \hat{r}_x(d) + 1$\Comment{Count requests}
\If{$d \in C(x)$}
\State \Return content to client
\EndIf
\If{$|C(x)| < B$}
\State Add $d$ to $C(x)$\Comment{On backward pass}
\ElsIf{$\hat{r}_x(d)\min_{y\in\mathcal{N}(x)} Q_y(d, y) > min_{d'\in C(x)}\hat{r}_x(d')\min_{y\in\mathcal{N}(x)} Q_y(d', y)$}
\State Replace lowest scoring item in $C(x)$ with $d$\Comment{On backward pass}
\EndIf
\State $y \gets \min_{y\in\mathcal{N}(x)} Q_x(d, y)$\Comment{Routing decision}
\State Route request for $d$ to $y$, measuring $t$
\State Receive $q = \min_{y'\in\mathcal{N}(y)} Q_y(d, y')$
\State $Q_x(d,y) \gets Q_x(d,y) + \alpha\left(t + q - Q_x(d,y)\right)$
\EndProcedure
\end{algorithmic}
\end{algorithm}

Algorithm~\ref{alg:qrouting} describes the behavior of a single node. The request distribution is
approximated by request counting, on line 4. Then, the caching decision is made. On the
backward path, if the cache is not full, the requested item is  added (line 8).
Otherwise, it is only placed if its expected cost is higher than the lowest scoring cache item
(line 10). The routing decision is taken in line 11, and the minimum in Equation~\eqref{eq:update} is
returned on acknowledgement (line 13). Finally, the Q table is adjusted according to
Equation~\eqref{eq:update}.

\section{Cost Minimization}

\label{sec:local1}

In this section we relate Q-caching to the cost minimization problem.  We introduce the local optimization problem in Section~\ref{sec:local}, and then use it to relate Q-caching to other policies in Section~\ref{sec:localrelated}.

\subsection{Local Optimization Problem}

\label{sec:local}
In this section, we consider the local optimization problem faced by each cache-router.

\begin{footnotesize}
\begin{table}[h!]
\center
\footnotesize
\begin{tabular}{ll}
\hline
\footnotesize
variable & description \\
\hline
$h_x(d)$ & hit probability \\
$r_x(d)$ & probability that request is for content $d$ \\
$\lambda_x$ & total request arrival rate (requests/s) \\
$\lambda_x(d)$ &  request arrival rate for content $d$, $\lambda_x(d) = \lambda r_x(d)$ (requests/s) \\
$E[D_x]$ & expected delay  experienced by node $x$ \\
$Q_x(d,y)$ & cost to go (computed using Q-routing) \\
$Q_x(d)$ & minimum cost to go   (computed using Q-routing) \\
$B$ & buffer size \\
$w_x(d)$ & weight equal to $r_x(d) Q_x(d)$ \\
\hline
\end{tabular}
\caption{Notation}  \label{tab:notation}
\end{table}
\end{footnotesize}

All quantities in this section account for a single tagged cache $x$.
Let $\lambda_x$ be the request arrival  rate.  Let $\lambda_x(d)$ be the request
rate for content $d$.  Then, $\lambda_x(d) = \lambda_x r_x(d)$, where $r_x(d)$ is the fraction of requests for content $d$.  
Let $h_x(d)$ be the hit probability of content $d$ at cache $x$, i.e., $h_x(d)$ is the probability that
 content $d$ is stored at cache $x$.   
The average delay to download a typical content at cache $x$ is $E[D_x] = \sum_d r_x(d) (1-h_x(d)) Q_x(d)$.  A summary of the notation used throughout this paper is presented in Table~\ref{tab:notation}.

%
%
The optimization problem faced at  cache-router $x$ is,
\begin{eqnarray}
\max && \sum_d r_x(d)  h_x(d) Q_x(d) \label{eq:opt1}  \\
&& \sum_d h_x(d) = B \label{eq:opt2}
\end{eqnarray}
Note that, as in~\cite{bianchi2013general},  we consider an expected buffer size constraint, i.e., the number 
of expected items in the cache cannot exceed the buffer size $B$.  

Let  $w_x(d) = r_x(d)  Q_x(d)$. 
The weight $w_x(d)$ is approximated online as the product of the number of request counts to a content multiplied by the 
cost to go.
%
%
The optimal solution to the problem above consists of storing the contents that have larger values of $w_x(d)$, i.e., setting $h_x(d)=1$ for the contents with larger values of $w_x(d)$ and $h_x(d)=0$ otherwise.  This  motivates the caching policy proposed in this paper,  which consists of storing the items with largest weights (i.e., largest costs  $w_x(d)$)  and evicting those with \emph{minimum expected cost}  (MEC).

\subsection{Contrasting MEC Against Other Policies}

\label{sec:localrelated}

The optimal solution of~\eqref{eq:opt1}-\eqref{eq:opt2} provides insight about how MEC compares against pre-existing policies.  If $Q_x(d)=1$, for all $d$, we note that MEC  degenerates to storing only the most popular items. This intuitive policy has formally shown to be optimal in~\cite{liu1997static}.  

Consider now the case where the popularity of the items is not known beforehand.  In that setup, a  cache which does not take into account the topology and costs of the remainder of the network when making its eviction decision can use LRU or LFU as proxies to the strategy of storing only the most popular items.    

MEC, in contrast,  
 accounts for the topology of the remainder of the network when making its eviction decisions.  While doing so, each cache 1) assumes that the network is optimized and 2) does not account for the impact of local decisions on the state of the other nodes in the network.  Together, properties 1) and 2) provide a simple and distributed strategy, which however is suboptimal in general.  The optimal routing and placement problem, accounting for all cache interddependencies, can be shown to be NP-hard~\cite{neves2010solving}.

\section{Q-caching Features} \label{sec:features}

Next, we discuss some of the features derived by combining Q-routing with MEC.

\subsection{Utility-Driven Caching}

Q-routing naturally allows for utility-driven routing.  By leveraging the cost-to-go computed by Q-routing to drive caching decisions, we also allow for utility-driven decisions  for content storage.  In particular, this provides additional flexibility with respect to existing policies such as LFU or LRU, whose utility is essentially  coupled to the content hit rate.

Note that LRU automatically stores all content that passes through a given cache.  Under Q-caching, in contrast, the controller might decide not to store  a given content which passes through the cache.    This, in turn, may lead to less churn.

\subsection{Enabling Content Diversity}

Q-caching promotes content diversity in the system, by favoring the storage of different contents at different caches.   Under LRU, the most frequently requested contents are always locally stored, irrespectively of the state of their neighbors.  The state of the neighbors affects a cache only through their miss streams.  Under Q-caching, in contrast, the state of the neighbors directly impacts the caching decisions at a given tagged cache.  This is because the cost-to-go is distributedly computed taking into account the distribution of the content in the network. One of the consequences of increased content diversity is reduced load at the custodians, in addition to reduced expected download times.

\subsection{Handling Loops}

One of the critical aspects of any routing algorithm relates to the handling of loops. 
Initial work on  ICN   suggested that  Pending Interest Tables (PITs)  could prevent loops \cite{Dai:2012,jacobson2009networking}.   This is because  PITs   associate to each pending request  a  random \emph{nonce}, which can be used to discard duplicate requests.  
However,  recent work \cite{garcia2015enabling,garcia2015fault}  indicates that ICNs are subject to routing loop problems even when cache-routers are equipped with PITs.


Routing in ICN can rely on Distance-Vector protocols, like Bellman-Ford, or Link-State Protocols, like Dijkstra. While the Dijkstra algorithm gracefully handles positive loops, Bellman-Ford based algorithms may lead to infinite relaying under adverse conditions.
As Q-routing is based on Bellman-Ford, it is subject to the \emph{count-to-infinity} problem, i.e., the convergence might take significant time in the worst case. 

The \emph{count-to-infinity} problem is exacerbated by the continuing variation of content availability due to caching. In this paper, we deal with this by not making the Q updates immediately influence the routing decisions. Instead, the Q-table used for routing is updated only periodically, allowing the Q values to settle in between routing updates. Using such a separate ``target" Q-table has proven to be a stabilizing factor in recent work on large-scale reinforcement learning~\cite{mnih_human-level_2015}. We leave a more theoretical study of the convergence to future work.

\section{Workload}
\label{workload}

Next, we present the workload used in our trace-driven numerical evaluations. 
 The physical topology is obtained from the Brazilian National Research Network (RNP).
\footnote{The topology is obtained from \url{http://web.archive.org/web/20140905111428/http://www.rnp.br/servicos/conectividade/rede-ipe}}   
The number of clients at each point of presence (PoP) is assumed to be proportional to the number of students of that region given by the 2013 university students census \cite{inep}.  

\begin{figure}
\centering
\includegraphics[scale=0.6]{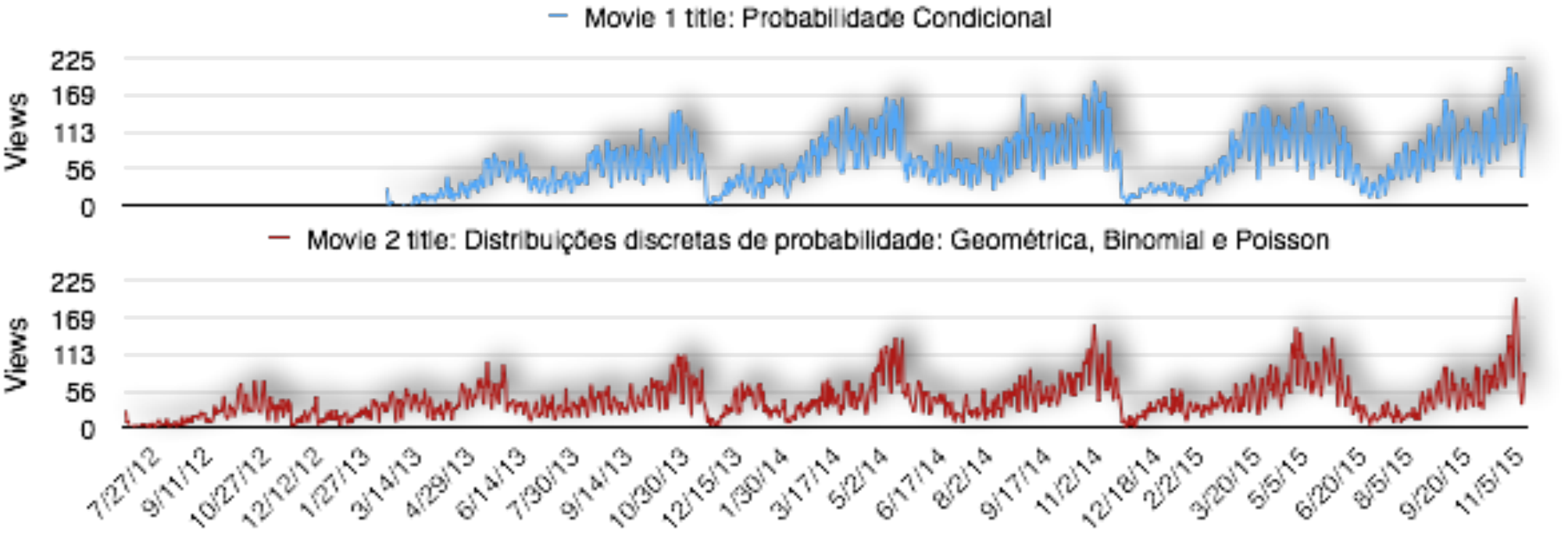}
\vspace{-0.1in}
\caption{Time series associated to  the top two most popular movies}
\label{fig:timeseries} \label{fig:views_evolution}
\vspace{-0.2in}
\end{figure}

We consider a content catalog comprising 31 Youtube videos from the UFRJ Performance Evaluation channel. 
The videos were published between 2012 and 2015, and count with up to 200,000 views and up to 700 followers.  
 Figure~\ref{fig:views_evolution} illustrates the time series of the two most popular movies. Each series start at the movie 
 publishing date. 
  As a side contribution of this work, we make available the  collected traces~\cite{traces}.  
Among the interesting properties extracted from the data, we point out the following (Figure~\ref{fig:views_evolution}):

\textbf{Periodicity:}  We note the periodic behavior of the number of views.  During the semester, the number of views grows.
  During vacations, it decreases.  During working days, the number of views is usually smaller than during the weekends.
  The trend is consistent among the movies.


\textbf{Non-stationarity: } As a general trend, the number of views increases from an year to the other.  
    In our trace-driven simulations we consider non-stationary workloads.  The simulation is divided into epochs, where
    each epoch corresponds to a day in the trace.
    The arrival rate for each content at each day is assumed to be proportional 
    to the number of requests issued for that content, which is obtained from the trace.

\begin{figure}
\centering
\includegraphics[scale=0.6]{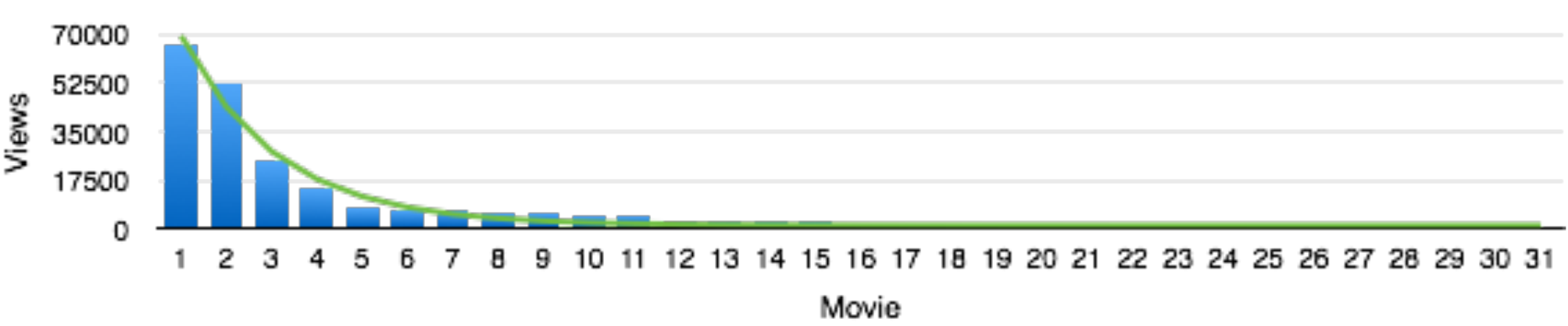}
\vspace{-0.1in}
\caption{Distribution of the number of views per movie}
\label{fig:distribution}
\vspace{-0.2in}
\end{figure}

Figure~\ref{fig:distribution} shows the distribution of the accumulated number of views by November of 2015.
The bars correspond to the collected data, and the line is  result of curve fitting.        
The distribution popularity is well approximated by an exponential distribution, with $y=1473+108947 \exp(-0.4707 x)$. 
The root mean square of the residuals  equals 2393. 
 As a comparison, the minimum root mean square of the residuals obtained with a power law distribution equals 4024. 

\section{Experiments}
\label{sec:experiments}

Next, we report the numerical results of our experiments. 
 We first consider simple topologies in Section~\ref{sec:simpletop}.
   Then, we contrast different solutions under the RNP topology in Section~\ref{rnpresults}. The major  strategy combinations considered are shown 
   in Table~\ref{tab:strategies}.  We consider three caching strategies (LRU, LFU and MEC)
    and  two routing strategies (shortest path first routing (SPF) 
and Q-routing).  Under SPF, 
requests   take the   path with minimum cost towards the custodian, assumed to be fixed and given. 
In all cases, if a replica is opportunistically found,
 the request is immediately  served.



\setlength{\tabcolsep}{2pt} 

\begin{table}[h]
\centering
\footnotesize
\vspace{-0.1in}
\begin{tabular}{l||l|l|l}
\hline
 Algorithm & Routing & Caching & Notes  \\
 \hline
 \hline
Inform & Q-routing & LRU & LRU  is too volatile (high churn) to work with Q-routing \\  
Q-caching & Q-routing & MEC & less volatile than Inform, comparable to Q-LFU \\
Q-LFU  & Q-routing & LFU & comparable to MEC \\
SPF+LRU & shortest path first & LRU & see Section~\ref{rnpresults}  \\
\hline
\end{tabular}
\vspace{-0.1in}
\caption{Some of the  strategy combinations considered in this paper}
\label{tab:strategies} 
\vspace{-0.1in}
\end{table}

 In our simulations, run in Matlab, we assume stationary latency between hops,
  with zero download delays (ZDD assumption) on the backward path, which is assumed to be
   identical to the forward path. As such, when a piece of content is found, all caches on its
    request path are immediately updated in accordance to the ZDD assumption. 
    All nodes run in lock-step, and process their entire request queue on every step. The ability of
     Q-routing to dynamically route around congestion is therefore not used, but will be 
     the subject of future work with more detailed simulations.

\subsection{Simple Topologies}

\label{sec:simpletop}

\begin{figure}[h]
\begin{tabular}{c@{\hskip 0pt}c@{\hskip 0pt}c}
	\includegraphics[scale=0.22]{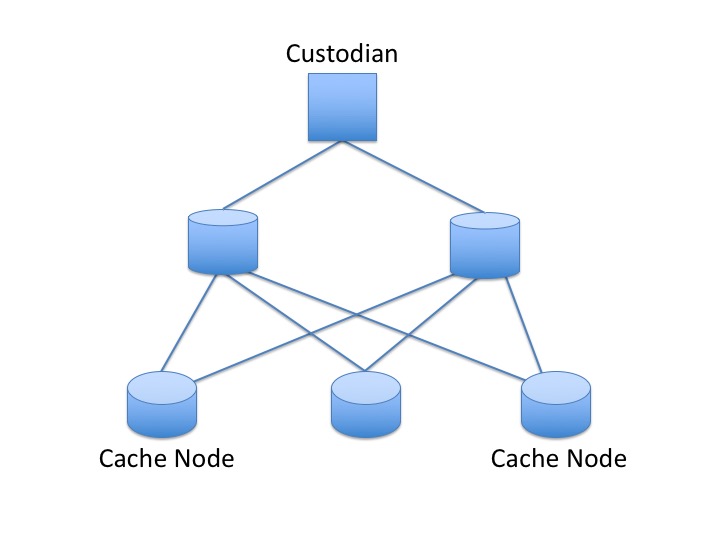} & 
\hspace{-0.3in}		\includegraphics[scale=0.22]{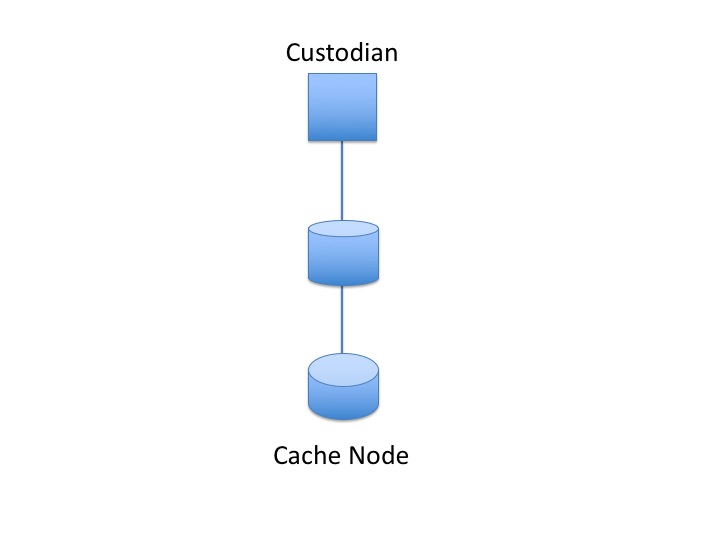} &
\hspace{-0.5in}	\includegraphics[scale=0.22]{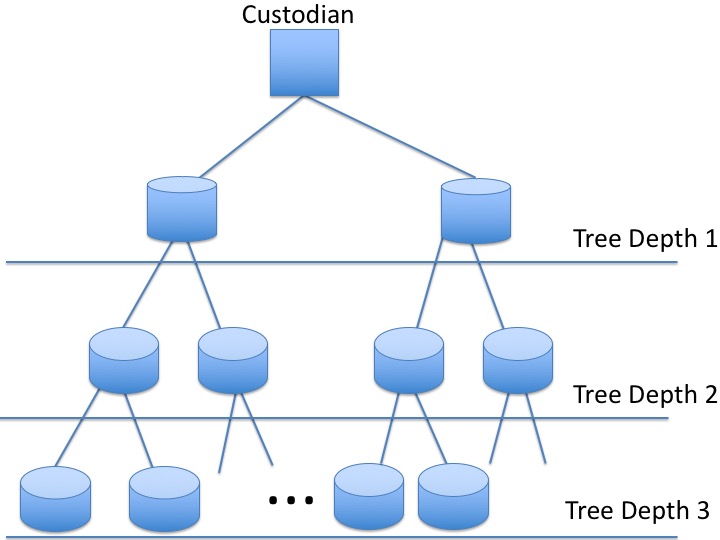} \\
(a) & \hspace{-0.3in}	(b) & \hspace{-0.5in}	(c) 
\end{tabular}
\vspace{-0.1in}
\caption{Hierarchical topologies: (a) layered; (b) chain and (c) tree. }
\label{fig:layered} \label{fig:chain}
\label{fig:treedepthtopology}
\end{figure}

Figure~\ref{fig:layered} presents the three simple topologies considered in our experiments.  
The layered topology (Figure~\ref{fig:layered}(a)) is inspired  by the Akamai network~\cite{sitaraman2014overlay}.  
The chain topology and the tree topology (Figures~\ref{fig:layered}(b) and (c)) are used to clarify the 
role of caching as opposed to routing in network performance.  Under the chain and tree topologies, 
all requests have a single path from the requester to the custodian, assumed to be in the root of the tree.

The reference setup considered  in our experiments is shown in Table \ref{tab:experiments}.  
The request arrival rate for the $k$-th content is given by $1/k^\alpha$, for $k=1, \ldots, C$, where $\alpha$ is referred to as the \emph{Zipf parameter}. 
Parameters are varied according to our experimental goals.  
For each parameter setting we simulated 10 runs to obtain the mean and the  95\% confidence interval 
of the metrics of interest, represented by
 a line and a shaded region in the plots that follow, respectively.  
  

\setlength{\tabcolsep}{0pt} 

\setlength{\tabcolsep}{2pt}

\begin{table}[h]
\centering
\footnotesize
\begin{tabular}{c|c|c|c|c|c|c}
\hline
  Number of nodes, & Cache & Zipf & Topology & Number of files, & Custodian & Exploration \\            
     $N$   &  size, $B$ & parameter, $\beta$ &  &  $C$ &cost & rate \\
  \hline
12 & 10 & 0.8 & layered & 100 & 100 &0.05 \\
 \hline
 \hline

\hline
\end{tabular}
\vspace{-0.1in}
\caption{Reference setup for experiments}
\label{tab:experiments}
\vspace{-0.2in}
\end{table}


 \subsubsection{Optimal Exploration Rate}
 \label{sub:exp}

 Next, we study the impact of the exploration rate on the metrics of interest.  To this aim,
 we consider a network of 12 nodes, and vary the exploration rate.  
 The other parameters are shown in Table~\ref{tab:experiments}.  
 
\begin{figure}[h]
\begin{center}
\begin{tabular}{cc}
    \includegraphics[width=0.48\textwidth]{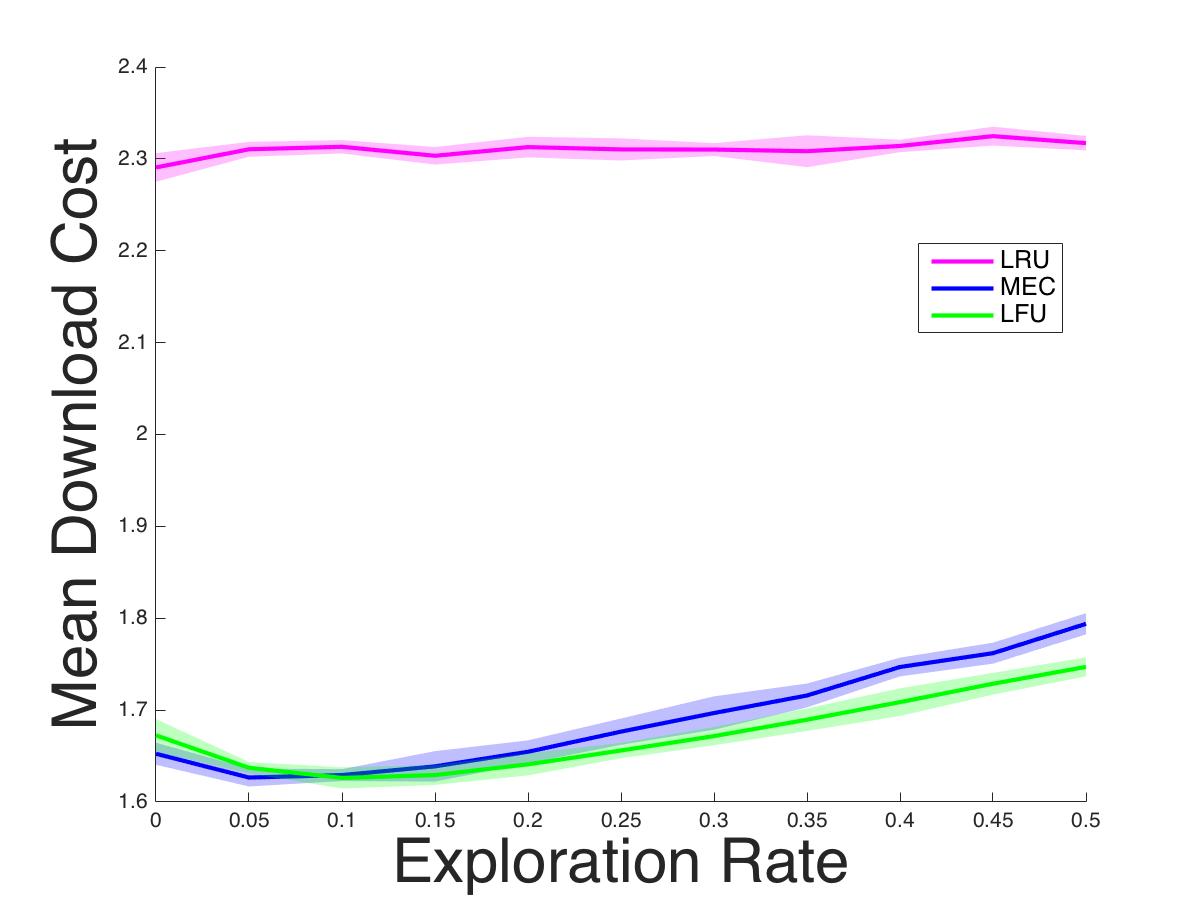}
& 
    \includegraphics[width=0.48\textwidth]{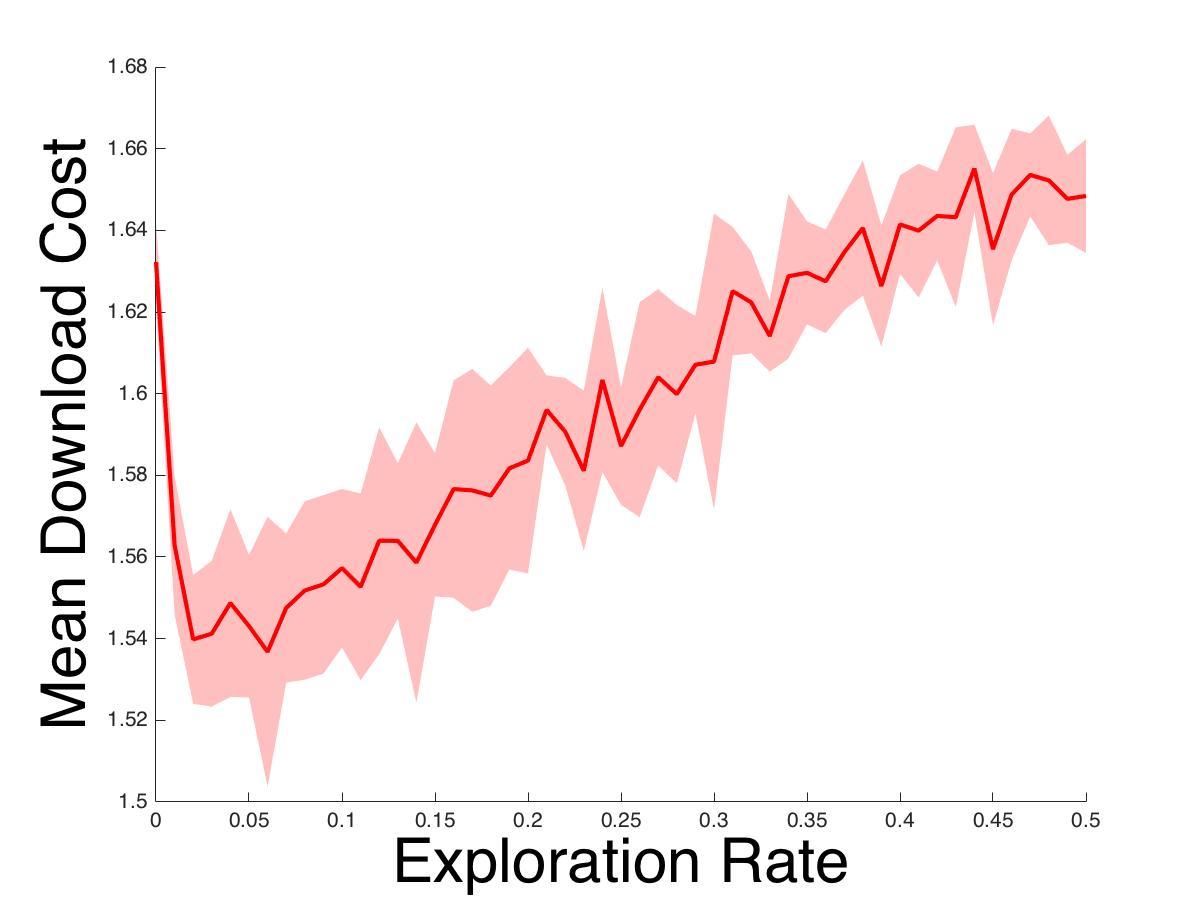} \\
    (a) three strategies & (b) MEC
    \end{tabular}
\end{center}
\vspace{-0.1in}
\caption{ Impact of exploration rate  }
\label{fig:exploration} \label{fig:exploration_onlymes}
\end{figure}

Figure \ref{fig:exploration}(a) shows how the mean download time varies as a function of the exploration rate.
 We consider a layered topology with  exogenous arrivals occuring at every node. 
Recall that exploration is necessary under MEC and Q-LFU in order to learn the distribution of cost-to-go.
Recall that the cost-to-go  is used for routing decisions by Q-routing and for caching decisions by MEC, i.e.,
when caching is performed using LFU or LRU, the cost-to-go is used only for routing decisions.  

 According to Figure~\ref{fig:exploration}(a), the optimal exploration rate is roughly 0.1 for MEC and 0.2 for Q-LFU.  
 Figure \ref{fig:exploration}(b) evaluates the mean download cost of MEC in more details, with a more fine grained set of exploration rates.
 The minimum mean download cost occurs when the exploration rate equals roughly 0.05.

Note that Q-LRU  
  is not sensitive to the exploration rate.
    We believe that the estimates of the cost-to-go are not very helpful under LRU due to the high content churn 
    associated to such a policy,
    which causes high variability in the distribution of items in the caches over time.  
    This variability  precludes the quick convergence and gains of Q-routing.


\subsubsection{MEC Increases Space Diversity}
\label{sub:space_diversity}

\begin{figure}[h]
\begin{center}
	\includegraphics[scale=0.2]{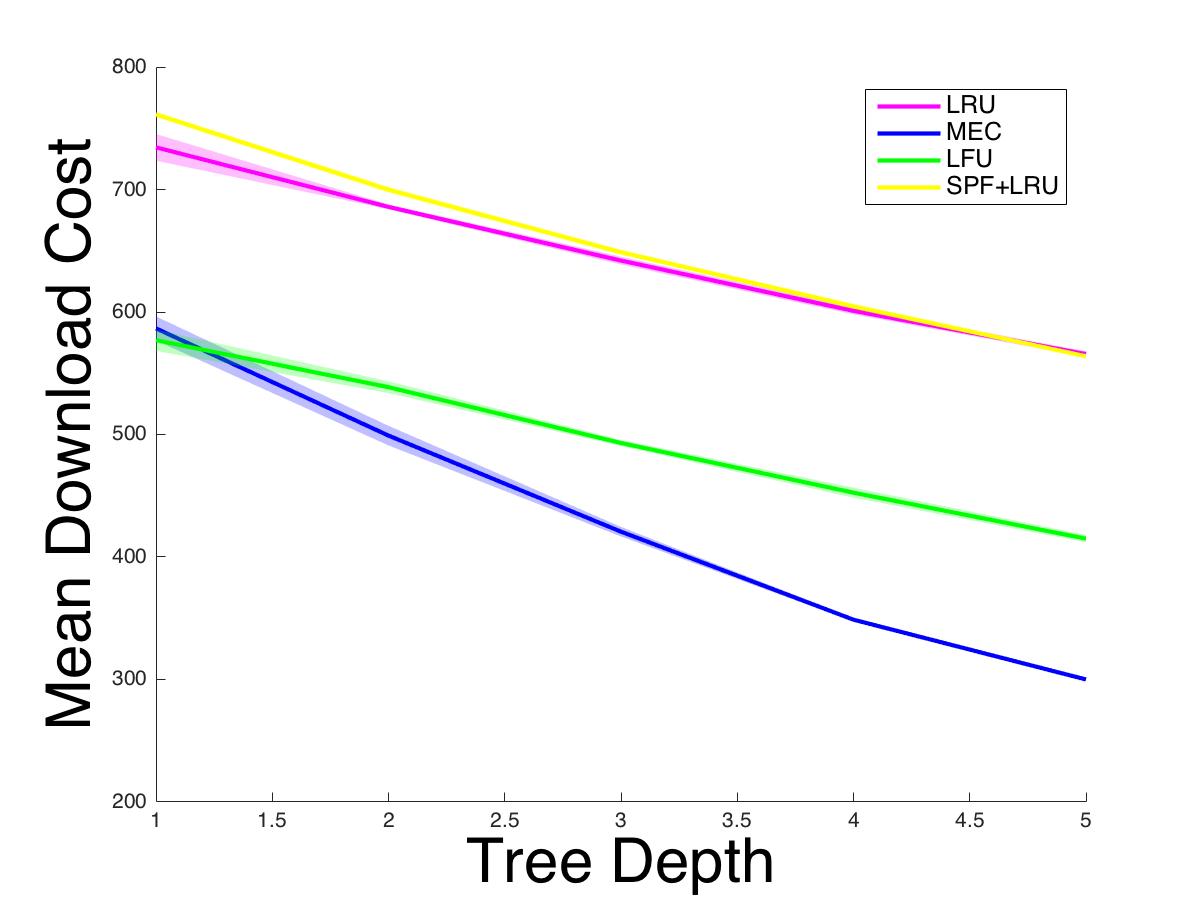}
\end{center}
\vspace{-0.2in}
\caption{Impact of tree depth on mean download time  }
\label{fig:treedepth}
\end{figure}

Next, we indicate that MEC increases space diversity, and as a consequence decreases mean download time.  
 To this aim, we consider a tree topology with $l$ levels and $2^{l+1}-1$ nodes.   The remaining  parameters are shown in Table~\ref{tab:experiments}.
Figure \ref{fig:treedepth} shows how the mean download time varies as a function fo the tree depth.   As the tree depth increases,  the mean download time decresaes more significantly under MEC as opposed to other strategies.  This indicates that MEC, by  exploring space diversity, stores different replicas of distinct items over the network, allowing requests to opportunistically download content from within the network and avoiding requests to the custodian, assumed to be associated to the most costly link in the system.

\subsubsection{MEC is Cost-aware}
\label{sub:cost_aware}

\begin{figure}[h]
\begin{center}
\begin{tabular}{cc}
    \includegraphics[width=0.48\textwidth]{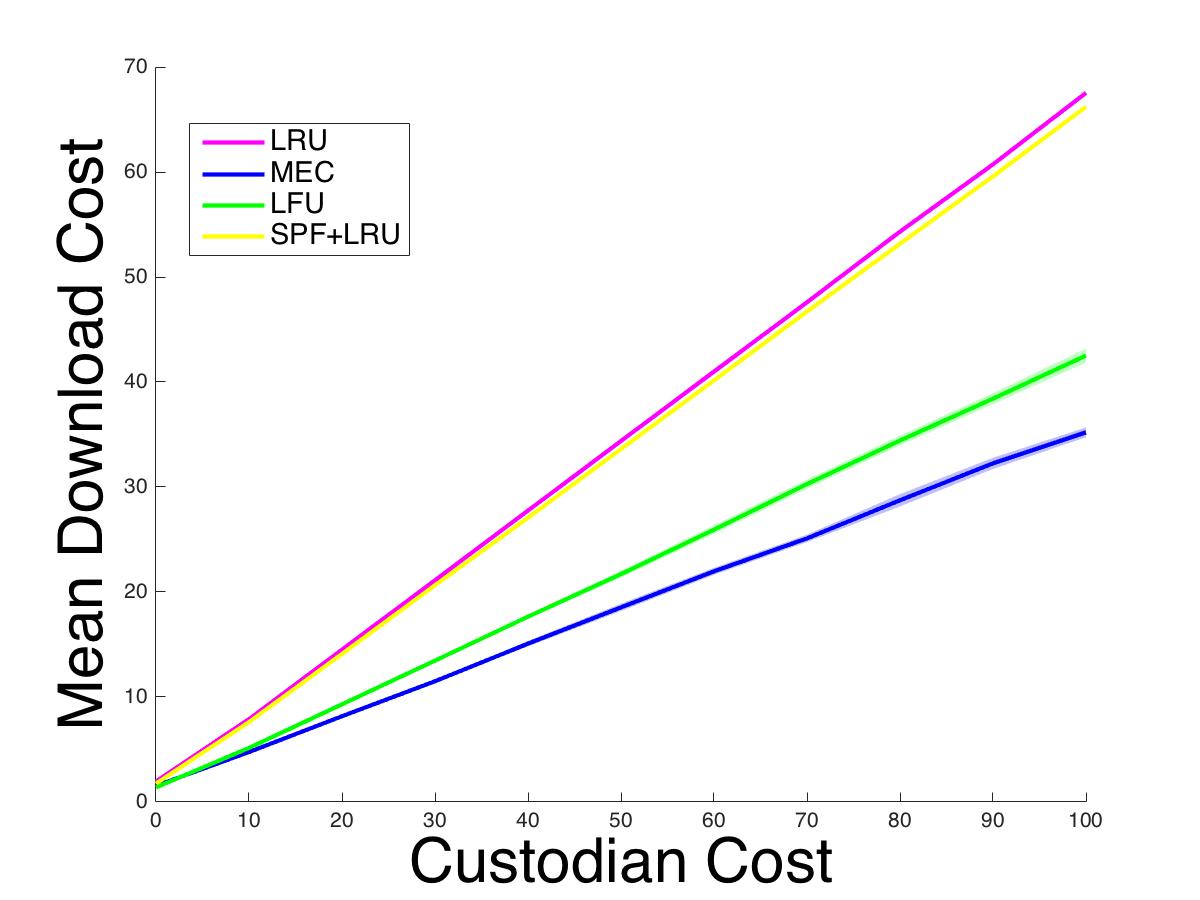}
& 
    \includegraphics[width=0.48\textwidth]{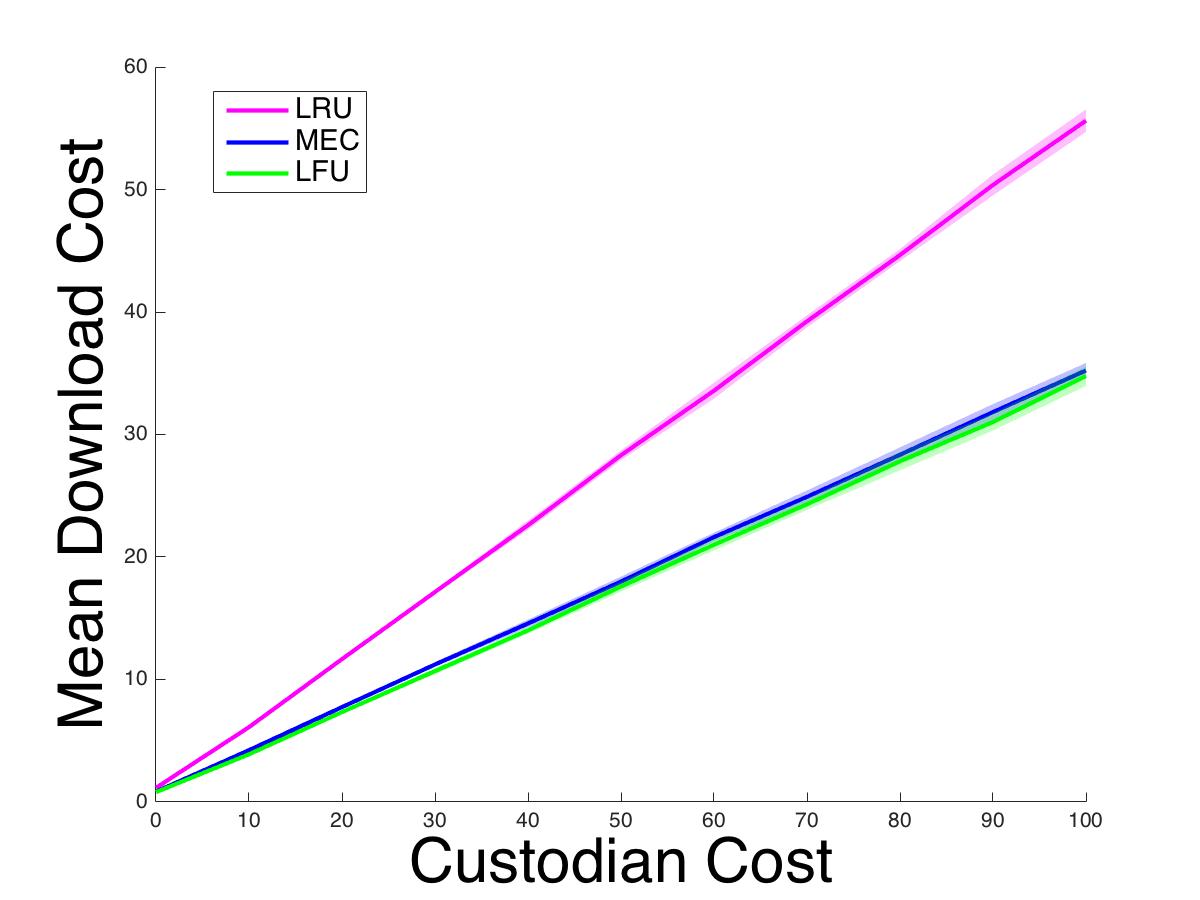} \\
    (a) & (b)
    \end{tabular}
\end{center}
\vspace{-0.1in}
\caption{ Impact of custodian cost }
\label{fig:custodian_cost} \label{fig:custodian_cost_serverhits}
\vspace{-0.3in}
\end{figure}

Figure \ref{fig:custodian_cost}(a) shows the impact of the costs of accesses to the  
custodian on the mean download time. We consider the reference setup of Table \label{tab:experiments}, with custodian cost varying between 0 and 100.  
Although the mean download time increases linearly with respect to the cost for all the strategies considered, the slope of  MEC is smaller than its counterparts.  This indicates that MEC, by being cost-aware, can gracefully handle changes in  custodian costs.  Figure~\ref{fig:custodian_cost}(b) shows how the number of server hits varies as a function of the custodian cost.  While MEC adjusts itself to account for cost changes, the other strategies considered are cost-oblivious and maintain the same server hit rate irrespective of server costs.

\subsubsection{Impact of Request  Locality}
\label{sub:locality}
\begin{figure}[h!]
\begin{center}
\begin{tabular}{cc}
    \includegraphics[width=0.45\textwidth]{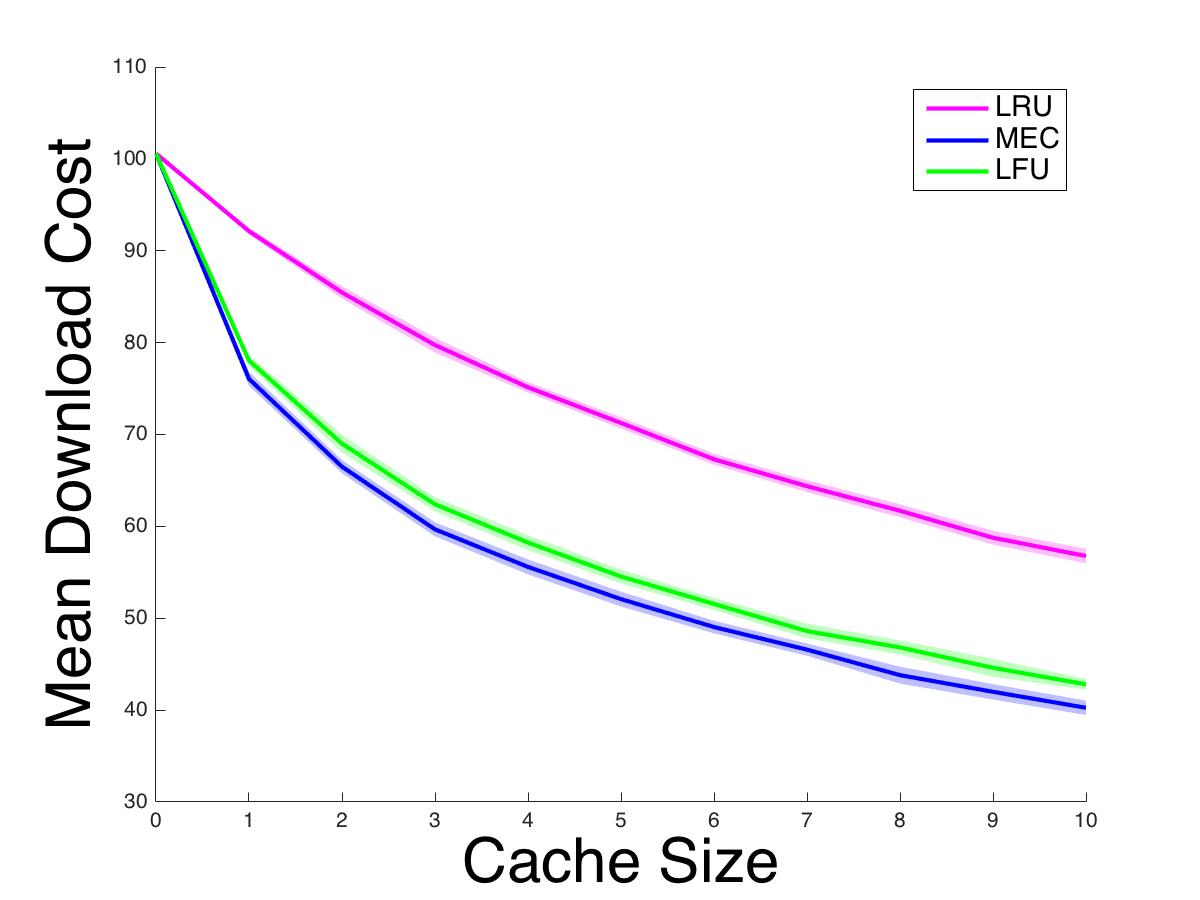}
& 
    \includegraphics[width=0.45\textwidth]{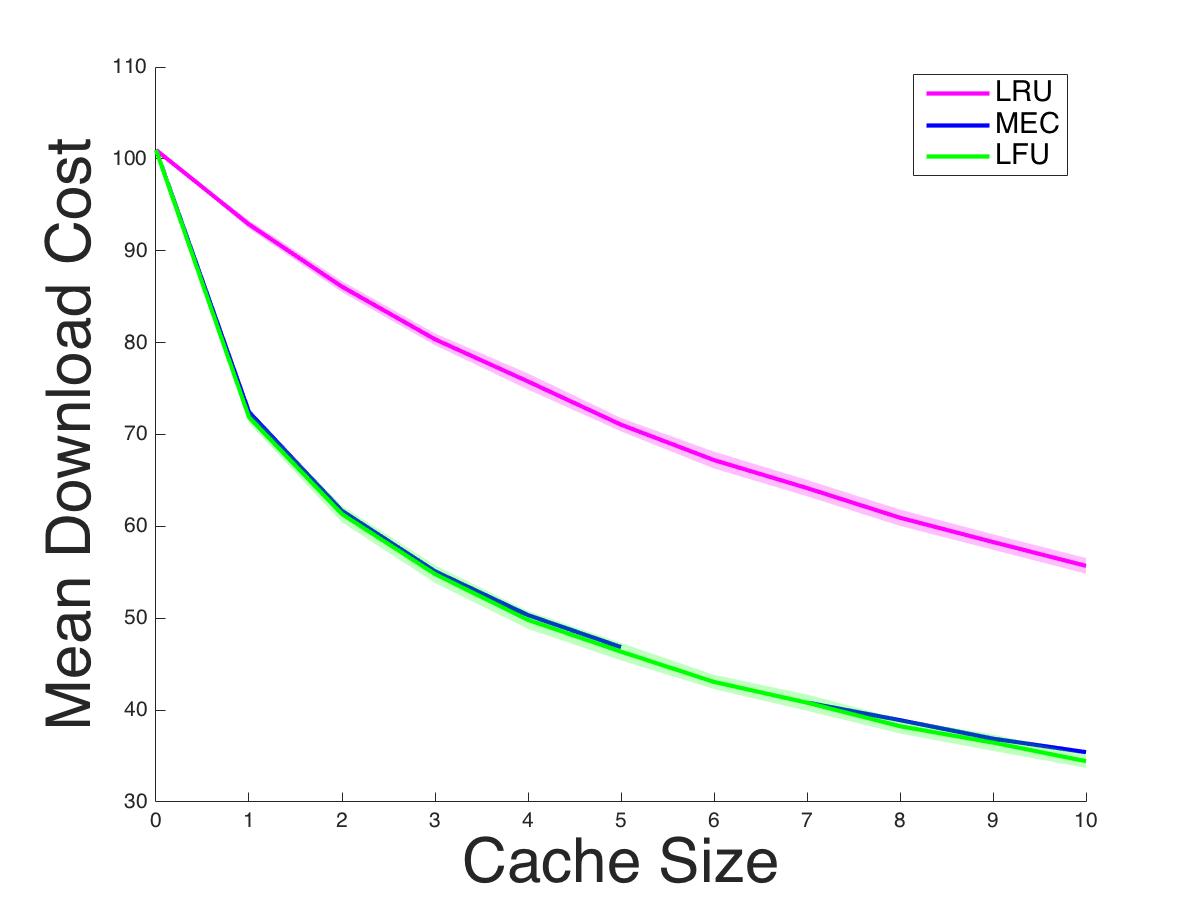} \\
    (a) arrivals  at every node & (b) arrivals only at the edge (leaves)
    \end{tabular}
\end{center}
\vspace{-0.1in}
\caption{Impact of request locality}
\label{fig:arrivals on every node} \label{fig:arrivals only on edges}
\end{figure}

Next, our goal is to study the impact of request locality on mean download time. 
We consider the reference setup shown in Table~\ref{tab:experiments}. 
%
%
 In Figures~\ref{fig:arrivals on every node}(a) and ~\ref{fig:arrivals on every node}(b) we consider the cases
  where exogenous requests arrive at every node and only at the leaves, respectively. 
As Figure \ref{fig:arrivals on every node}(a) shows, MEC is better that Q-LRU, and slight better than Q-LFU,  
when there are exogenous arrivals at every node.
 In contrast,  Figure \ref{fig:arrivals only on edges}(b) shows that LFU is slightly better than MEC when there are 
 exogenous arrivals only at the edge.    This is because, due to symmetry, the cost-to-go associated 
 to items which are not stored in the leaves  is roughly the same for  all nodes.  
 Therefore, the gains of MEC due to heterogenous cost-to-go values are not leveraged.

\subsection{RNP Topology and Youtube Traces}  \label{rnpresults}
To validate the strategies proposed,  real network topology and video traces are used as described in Section \ref{workload}. The video custodian  is placed at João Pessoa. Every point of presence of the  RNP network is associated to a cache-router with capacity to store up to three videos. All videos have unit size. The cost of a link is inversely proportional to its bandwidth.  We set 20 Gbps as  the reference bandwidth, meaning that its cost is normalized to 1.  Henceforth, we consider only normalized costs, referred to simply as \emph{costs}.   The custodian has a link with capacity of  100 Mbps, i.e., its cost is 200.   

\begin{figure}[h]
\begin{center}
\begin{tabular}{cc}
    \includegraphics[width=0.5\textwidth]{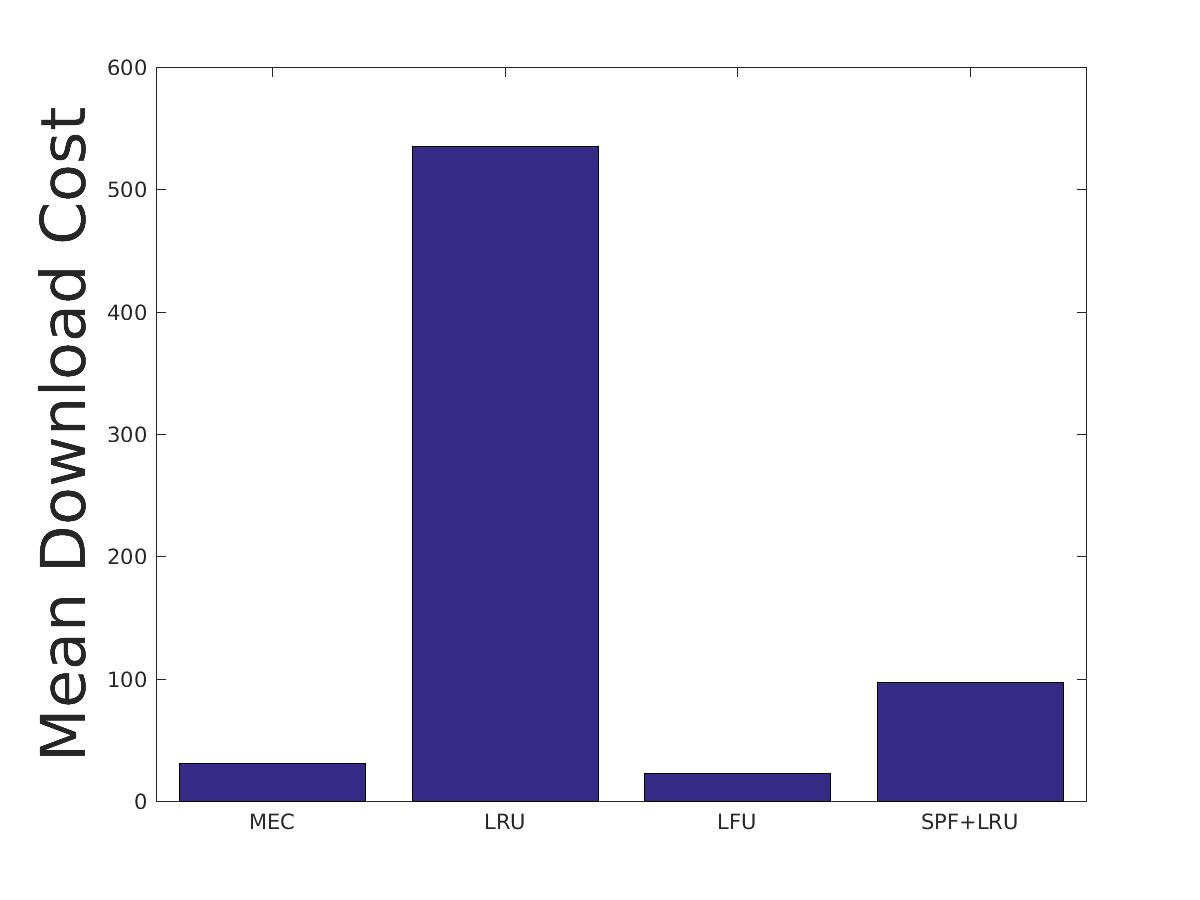}
& 
    \includegraphics[width=0.5\textwidth]{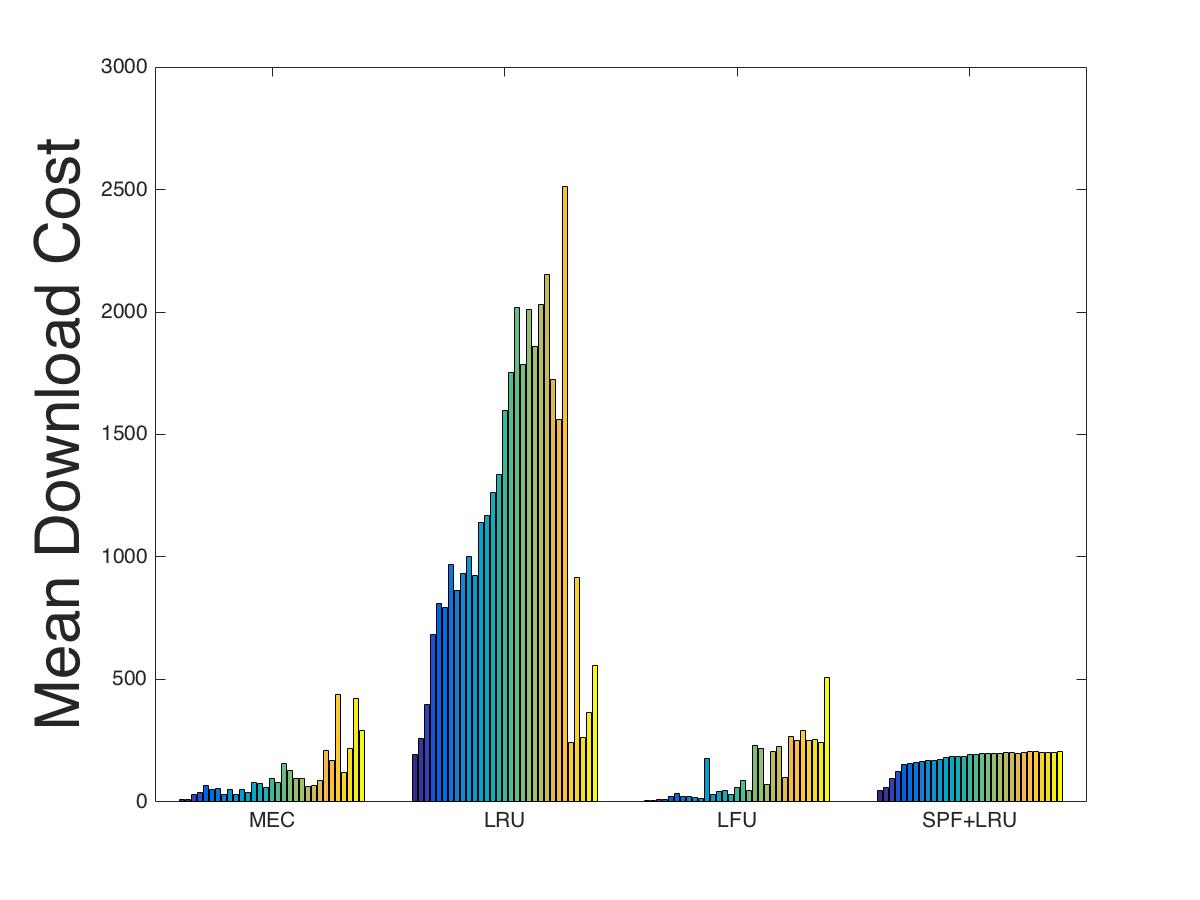} \\
    (a)  averaged over all files  & (b) per file 
    \end{tabular}
\end{center}
\vspace{-0.1in}
\caption{Mean download costs (RNP topology and Youtube video traces)}
\label{fig:rnpexp1}
\end{figure}

Figures ~\ref{fig:rnpexp1}(a) and (b) show the results obtained using the RNP topology and the Youtube trace-driven simulations. Figure~\ref{fig:rnpexp1}(a)  shows the  mean download cost to retrieve all files for the four considered strategies. As the figure indicates, Q-Routing+LFU is slightly better than Q-Routing+MEC, and Q-Routing+LRU is the worst strategy.
Figure~\ref{fig:rnpexp1}(b)  shows the  mean download cost for each video. As in Figure~\ref{fig:distribution},  videos are sorted according to the total number of views. Under Q-Routing+LRU,  the mean download cost is high for all videos. Under Q-Routing+MEC and Q-Routing+LFU, the popular videos have smaller minimum download costs.  Under MEC, 26 videos have mean download costs lower than 200, in agreement with  the hypothesis that MEC increases space diversity. Under LFU, in contrast, only  20 videos have mean download cost lower than 200. 
Under SPF+LRU,  the four most popular videos have    mean download costs lower  than 200. The mean download cost of all other videos is roughly equal to 200.


\begin{figure}[h]
\begin{center}
    \includegraphics[width=0.5\textwidth]{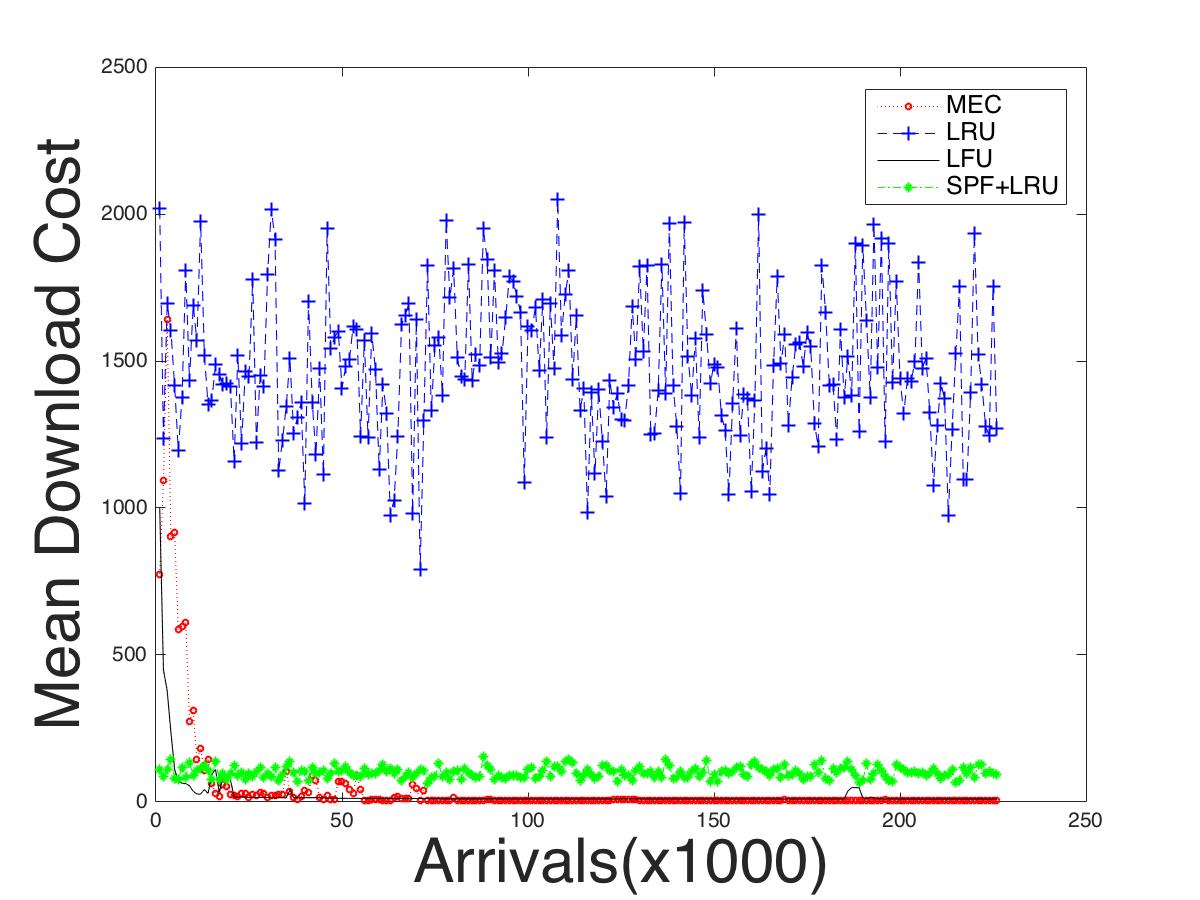}
\end{center}
\vspace{-0.2in}
\caption{Evolution of the mean download cost over time}
\label{fig:rnpexpdynamics}
\end{figure}

Figure~\ref{fig:rnpexpdynamics} shows the evolution of the mean download cost over time for the four considered strategies. This figure shows although the mean download cost of Q-Routing+LRU widely varied during the simulation period,  SPF+LRU showed a better and more stable behavior. As  previously discussed, this occurs because LRU yields high content   volatility in the caches, not allowing  Q-Routing to properly converge.
Under LRU, every new request to a content that is not cached  leads to an eviction, naturally producing more churn, which needs to be coped by Q-Routing. Under MEC and LFU, in contrast,   a popular content might  never show up in the miss stream. This feature promotes stabilization.  

Note that the convergence of MEC is slower than that of LFU.  Nonetheless,   after convergence the mean download costs of MEC and LFU are  4 and  10, respectively. 
 This means that depending on the dynamics of the content catalog and of the popularity of contents,    MEC might be preferred over  LFU.

\setlength{\tabcolsep}{6pt}

\section{Related Work}

\label{sec:related}

The literature on information centric networks, accounting for the relations between caching and routing~\cite{rossini2014coupling, chiocchetti2013inform, fayazbakhsh2013less}, analytical models~\cite{domingues2013enabling,  dabirmoghaddam2014understanding} and different naming schemes~\cite{bari2012survey} is rapidly growing.   To the best of our knowledge this paper is the first to leverage information provided by reinforcement-learning based routing algorithms for caching decisions.  


The role of routing and the interplay between routing and caching in ICNs is discussed in~\cite{yi2014role,chiocchetti2012exploit,chiocchetti2013inform}.  As such previous works suggest, jointly solving the routing and caching problems might lead to significant performance gains. However, the joint analysis of these two aspects is non-trivial, and taking them into account together in a simple manner was still an open problem.  In this paper, we propose to use the cost-to-go provided by Q-routing for caching decisions, integrating routing and caching in a simple way.

The evaluation of the interplay between caching and routing through analytical models has been recently 
subject of  considerable attention~\cite{rossini2014coupling}. 
  When determining the steady state occupancy of the caches, one approach consists of 
  searching for  a fix-point solution which is compatible with the routing and caching decisions \cite{rosensweig2010approximate}.  We believe that the solutions proposed in this paper are also amenable to analysis through adaptations of the models proposed in \cite{rosensweig2010approximate}.  

The use of reinforcement learning (RL) for decision making in ICNs has been considered in~\cite{bastosestrategia,chiocchetti2013inform}.    \cite{chiocchetti2012exploit} studied the classical RL tradeoff between exploitation and exploration  and then proposed Inform~\cite{chiocchetti2013inform}, which combines Q-routing with LRU for improved performance. \cite{bastosestrategia} also propose the use of RL  for routing decisions.  In this work, in contrast, we propose the use of RL for routing and caching decisions, in an integrated fashion.

\section{Discussion}
\label{sec:discussion}


Next, we discuss some of the simplifying assumptions considered in this work, together with possible directions for future work.

\textbf{Forward and Backward Routes:}  In this paper, we focused on the forward (upstream) routes, namely the routes between requestors and custodians.  We leveraged the opportunistic encounters between requests and replicas of the content closer to users.  The backward (downstream) routes were assumed to be the same as the forward routes, and the zero download delay (ZDD) assumption was considered.  If backward routes may be distinct from forward routes, additional flexibility might be gained to determine how to route the content to users, so as to strategically place new replicas where needed. 

\textbf{Routing and Caching Utilities:} We assumed that the utilities associated to routing and caching decisions are the same.  However, in general they might be associated to  distinct utilities. In such case, additional control messages might be required.  The gains, however, might compensate such overhead specially if routing costs (e.g., due to congestion) are very distinct from content costs  (e.g., due to  the availability of custodians for certain rare or unpopular content).

\section{Conclusion}
\label{sec:conclusion}

Caching and routing are two of the building blocks of ICNs.  In this paper, we showed how to leverage information provided by Q-routing in order to make caching decisions.  The proposed solution, Q-caching, is  simple and flexible, allowing for different utilities and metrics of interest to be taken into account when coupling caching and routing for increased performance.  We numerically investigated the effectiveness of Q-caching, and contrasted it against different schemes, including the state-of-the-art Inform (Q-routing coupled with LRU).  We have found that in the scenarios investigated Q-caching  is either better or competitive againts the considered counterparts.  Future work consists of establishing the conditions under which Q-caching is optimal, and   studying its converence properties.


\footnotesize

\bibliographystyle{sbc}

\bibliography{main}

\end{document}